\documentclass[reprint,aps,pre,floatfix,twocolumn,showpacs,superscriptaddress, nofootinbib
,longbibliography
]{revtex4-1}

\usepackage{times}
\usepackage[pdftex]{graphicx}
\usepackage{latexsym,amsmath}
\usepackage[english]{babel}
\usepackage[pdftex]{color}
\usepackage{microtype}
\usepackage{comment}
\usepackage[utf8]{inputenc}
\usepackage[normalem]{ulem}
\usepackage{etoolbox}
\usepackage[justification=centerlast,labelfont=bf,font=small]{caption}
\usepackage{lineno}
\usepackage[shortlabels]{enumitem}

\usepackage{latexsym,amsmath}
\usepackage{amsbsy}
\usepackage{amssymb}
\usepackage{epstopdf}
\usepackage{comment}
\usepackage{multirow,tabularx}
 \usepackage[normalem]{ulem}

\usepackage{color}

\begin{document}

\title{Collective response to the media coverage of COVID-19 Pandemic on Reddit and Wikipedia}

\author{Nicol\`o Gozzi}
\affiliation{Networks and Urban Systems Centre, University of Greenwich, London, UK}
\author{Michele Tizzani}
\affiliation{ISI Foundation, Turin, Italy}
\author{Michele Starnini}
\affiliation{ISI Foundation, Turin, Italy}
\author{Fabio Ciulla}
\affiliation{Quid Inc., San Francisco, USA}
\author{Daniela Paolotti}
\affiliation{ISI Foundation, Turin, Italy}
\author{Andr\'e Panisson}
\email{andre.panisson@isi.it}
\affiliation{ISI Foundation, Turin, Italy}
\author{Nicola Perra}
\email{n.perra@greenwich.ac.uk}
\affiliation{Networks and Urban Systems Centre, University of Greenwich, London, UK}

\date{\today}

\begin{abstract}

The exposure and consumption of information during epidemic outbreaks may alter risk perception, trigger behavioural changes, and ultimately affect the evolution of the disease. 
It is thus of the uttermost importance to map information dissemination by mainstream media outlets and public response.
However, our understanding of this exposure-response dynamic during COVID-19 pandemic is still limited. 
In this paper, we provide a characterization of media coverage and online collective attention to COVID-19 pandemic in four countries: Italy, United Kingdom, United States, and Canada. 
For this purpose, we collect an heterogeneous dataset including  $227,768$ online news articles and $13,448$ Youtube videos published by mainstream media, $107,898$ users posts and $3,829,309$ comments on the social media platform Reddit, and $278,456,892$ views to COVID-19 related Wikipedia pages. Our results show that public attention, quantified as users activity on Reddit and active searches on Wikipedia pages, is mainly driven by media coverage and declines rapidly, while news exposure and COVID-19 incidence remain high. 
Furthermore, by using an unsupervised, dynamical topic modeling approach, we show that while the attention dedicated to different topics by media and online users are in good accordance,
interesting deviations emerge in their temporal patterns.  
Overall, our findings offer an additional key to interpret public perception/response to the current global health emergency and raise questions about the effects of attention saturation on collective awareness, risk perception and thus on tendencies towards behavioural changes. 
 
\end{abstract}

\maketitle

\section{Introduction}

``In the next influenza pandemic, be it now or in the future, be the virus mild or virulent, the single most important weapon against the disease will be a vaccine. The second most important will be communication''~\cite{barry2009pandemics}. This evocative sentence was written in May 2009 by John M. Barry, in the early phases of what soon after become the H1N1 2009 pandemic. In his essay, Barry summarised the mishandling of the deadly 1918 Spanish flu highlighting the importance of precise, effective and honest information in the onset of health crises.

Eleven years later we find ourselves dealing with another pandemic.  The cause is not a novel strain of influenza, but these words are, unfortunately, still extremely relevant. In fact, as the SARS-CoV-2 sweeps the world and the vaccine is just a far vision of hope, the most important weapons to reduce the burden of the disease are non-pharmaceutical interventions~\cite{funk2010review,verelst}.
Social distancing became paramount, gatherings have been cancelled, mobility within and across countries have been dramatically reduced. While such measures have been enforced to different extents across nations, they all rely on compliance. Their effectiveness is linked to risk and susceptibility perception~\cite{hbm}, thus the information that citizens are exposed to is fundamental.

History repeats itself and we seem not be able to learn from our past mistakes. As happened in 1918, despite early evidences from China~\cite{wu2020estimating,WHO_timeline}, the virus was first equated, by many, to the normal seasonal flu. As happened in 1918, many national and regional governments organised campaigns aimed at boosting social activities (and thus local economies) actively trying to convince people that their cities were safe and that the spreading was isolated in far away locations. For example, the hashtag \#MilanoNonSiFerma (Milan does not stop) was coined to invite citizens in Milan to go out and live normally.  Free aperitifs were offered in Venice. In hindsight, of course, is easy to criticise the initial response in Italy. In fact, the country has been one of the first to experience rapid growth of hospitalizations~\cite{WHO_reports}. However, the Mayor of London, twelve days before the national lockdown, and few days after the extension of the cordon sanitaire to the entire country in Italy, affirmed via his official Facebook page ``we should carry on doing what we’ve been doing”~\cite{mayor_of_london}. More in general, in several western countries, the news coming from others reporting worrying epidemic outbreaks were not considered as relevant for the internal situation. This initial phase aimed at conveying low local risk and boosting confidence about national safety has been repeated, at different times, across countries. 
A series of surveys conducted in late February provide a glimpse of the possible effects of these approaches. 
They report that citizens of several European countries, despite the grim news coming from Asia, were overly optimistic about the health emergency placing their risk of infections to be $1\%$ or less~\cite{raude2020people}. As happened in 1918, the countries that reacted earlier rather than later were able to control the virus with significant less victims~\cite{kraemer2020effect,Maier742,anderson2020will,bedford2020covid,colbourn2020covid}.

History repeats itself, but the context often is radically different. In 1918, news circulated slowly via news papers, controlled by editorial choices, and of course words of mouth.
In 2009, we witnessed the first pandemic in the social media era. 
Newspapers and TV were still very important source of information, but Twitter, Facebook, YouTube, Wikipedia started to become relevant for decentralized news consumption, boosting peer discussions, and misinformation spread. 
Today these platforms and websites are far more popular, integral part of society and instrumental pieces of the national and international news circulations. 
Together with traditional news media, they are the principal sources of information for the public. 
As such, they are fundamental drivers of people perception, opinions, and thus behaviours. 
This is particularly relevant for health issues. For example,  about $60\%$ of adults in the USA consulted online sources to gather health information~\cite{fox2013health}. 

With respect to past epidemics and pandemics, studies on traditional news coverage of the 2009 H1N1 pandemic highlighted the importance of framing and its effect on people's perception, behaviours (such as vaccination intent), stigmatisation of cultures at the epicentre of the outbreak, and how these factors differ across countries/cultures~\cite{lee2014predictors,mccauley2013h1n1,lin2013effects,lee2013press,jung2012attention, keramarou2011two}. 
During Zika epidemic in 2016, public attention was synchronised across US states, driven by news coverage about the outbreak and independently of the real local risk of infection~\cite{tizzoni2020impact}.
With respect to COVID-19 pandemic itself, a recent study clearly shows how Google searches for ``coronavirus'' in the USA spiked significantly right after the announcement of the first confirmed case in each state~\cite{bento2020evidence}. 
Several studies based on Twitter data also highlight how misinformation and low quality information about COVID-19, although overall limited, spread before the local outbreak and rapidly took off once the local epidemic started.
In the current landscape, this has the potential to boost irrational, unscientific, and dangerous behaviours~\cite{gallotti2020assessing,singh2020first,cinelli2020covid}.

On the other hand, despite some important limitations~\cite{lazer2014parable}, modern media has become a key data source to observe and monitor health. In fact, posts on Twitter~\cite{culotta2010towards,lampos2010tracking,zhang2017forecasting,de2013predicting,de2013social,broniatowski2013national}, Facebook~\cite{araujo2017using}, and Reddit~\cite{park2017tracking,kumar2015detecting}, page views in Wikipedia~\cite{generous2014global,hickmann2015forecasting} and searches on Google~\cite{ginsberg2009detecting,dugas2013influenza} have been used to study, nowcast and predict the spreading of infectious diseases as well as the prevalence of noncommunicable illnesses. 
Therefore, in the current full-fledged digital society, information is not only key to inform people's behaviour but can be used to develop an unprecedented understanding of such behaviours, as well as of the phenomena driving them. 

The context where COVID-19 is unfolding is thus very heterogeneous and complex. 
Traditional and social media are integral parts of our perception and opinions, have the potential to trigger behaviour change and thus influence the pandemic spreading.
Such complex landscape must be characterized in order to understand the public attention and response to media coverage.
Here, we tackle this challenge by assembling an heterogeneous dataset which includes $227,768$ news and $13,448$ YouTube videos published by traditional media, $278,456,892$ views of topical Wikipedia pages, $107,898$ submissions and $3,829,309$ comments from $417,541$ distinct users on Reddit, as well as epidemic data in four different countries: Italy, United Kingdom, United States, and Canada.

First, we explore how media coverage and epidemic progression influence public attention and response. To achieve this, we analyze news volume and COVID-19 incidence with respect to Wikipedia page views volume and Reddit comments. Our results show that public attention and response are mostly driven by media coverage rather than disease spreading. Furthermore, we observe typical saturation and memory effects of public collective attention. Moreover, using an unsupervised topic modeling approach, we explore the different topics framed in traditional media and in Reddit discussions. We show that, while attentions of news outlets and online users towards different topics are in good accordance, interesting deviations emerge in their temporal patterns.  Also, we highlight that, at the end of our observation period, general interest grows towards topics about the resumption of activities after lockdown, the search for a vaccine against Sars-Cov-2, acquired immunity and antibodies tests.\\
Overall, the research presented here offers insights to interpret public perception/response to the current global health emergency, raises interrogatives about the effects of attention saturation on collective awareness, risk perception and thus on tendencies towards behavioural changes.

\section{Impact of media coverage and epidemic progression on collective attention}
\label{sec:1}
How is collective attention shaped by news media coverage and epidemic progression? To tackle this important question, we collected an heterogeneous dataset that includes COVID-19 related news articles and Youtube videos published online by mainstream information media, relevant posts and relative discussion of geolocalized Reddit users, and country-specific views to Wikipedia pages related to COVID-19 for Italy, United Kingdom, United States and Canada (see subsections A, B, C of Methods and Materials for details). This choice aims to provide an overview of media coverage and a proxy of public attention and response. On the one hand, the study of news articles and videos allows us to estimate the exposure of the public to COVID-19 pandemic in traditional news media. On the other hand, the study of users discussions and response on social media (through Reddit) and information seeking (through Wikipedia page views) allows us to quantify the reaction of individuals to both the COVID-19 pandemic and news exposure. As mentioned in the introduction, previous studies showed the usefulness of social media, internet use and search trends to analyze health-related information streams and monitor public reaction to infectious diseases \cite{eysenbach2011infodemiology, milinovich2014internetsurveillance, park2020covid19twitter, park2018reddithealth, lamb2012twitterpi}. 

Hence, we consider volume of comments of geolocalized users on the subreddit /r/Coronavirus\footnote{Subreddits are user-created areas of interest where discussions are organized in Reddit. /r/Coronavirus is the subreddit related to discussions surrounding COVID-19 pandemic} to explore the public discussion in reaction to media covering the epidemic in the various countries, while we consider the number of views of relevant Wikipedia pages about COVID-19 pandemic to quantify users interest. 
It is important to stress how Reddit and Wikipedia provide different aspects of online users behaviour and collective response.
In fact, while Reddit posts can be regarded as a general indicator of the online discussion surrounding the global health emergency, the number of access to COVID-19 related Wikipedia pages is a proxy of health information seeking behaviour (HISB). HISB is the act through which individuals retrieve and acquire new knowledge about a specific topic related to health \cite{nehama2017infoseeking, lambert2007hisb}, and it is likely to be triggered on a population scale by a disrupting event, such as the threaten of a previously unknown disease \cite{infoseekingh1n1, Pang2014CrisisbasedIS}. 

\begin{figure}[tbp]
\includegraphics[width=0.99\linewidth]{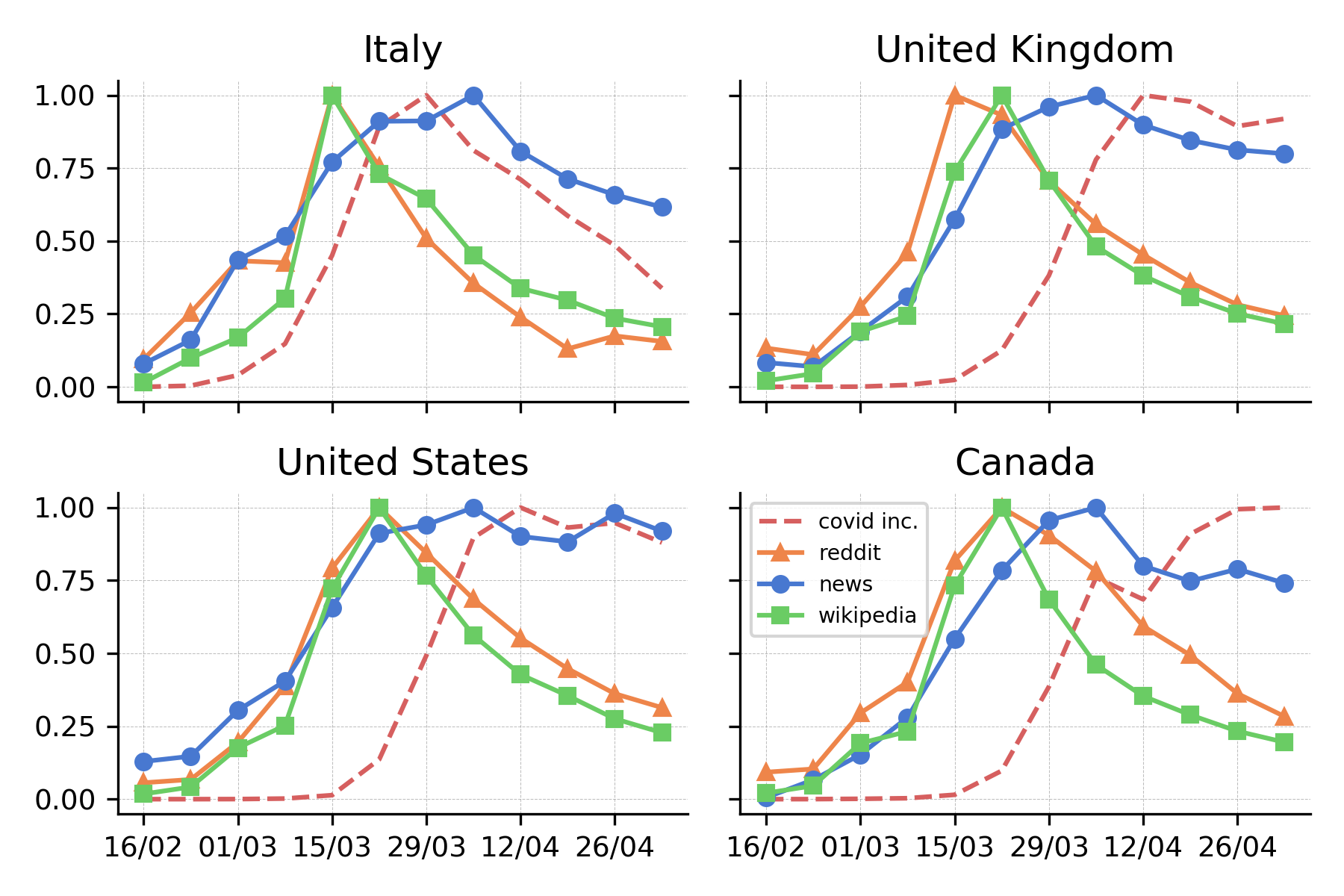} 
\caption{\label{fig:volume} Normalized weekly volume of news articles and Youtube videos (news)), Reddit comments (reddit), Wikipedia views (wikipedia) related to COVID-19 pandemic and COVID-19 incidence (covid inc.) in different countries.}
\end{figure}

Our analysis starts by comparing, in Figure~\ref{fig:volume}, the weekly volume of news and videos published on Youtube, Wikipedia views, and Reddit comments of geolocalized users in comparison with the weekly COVID-19 incidence in the four countries considered. It can be seen how, as COVID-19 spreads, both media coverage and public interest grow in time. However, public attention, quantified by the number of Reddit comments and Wikipedia views, sharply decreases after reaching a peak, despite the volume of news and COVID-19 incidence remaining high. Furthermore, the peak in public attention consistently anticipates the maximum media exposure and maximum COVID-19 incidence. 

\begin{figure}[tbp]
\centering
\includegraphics[width=0.99\linewidth]{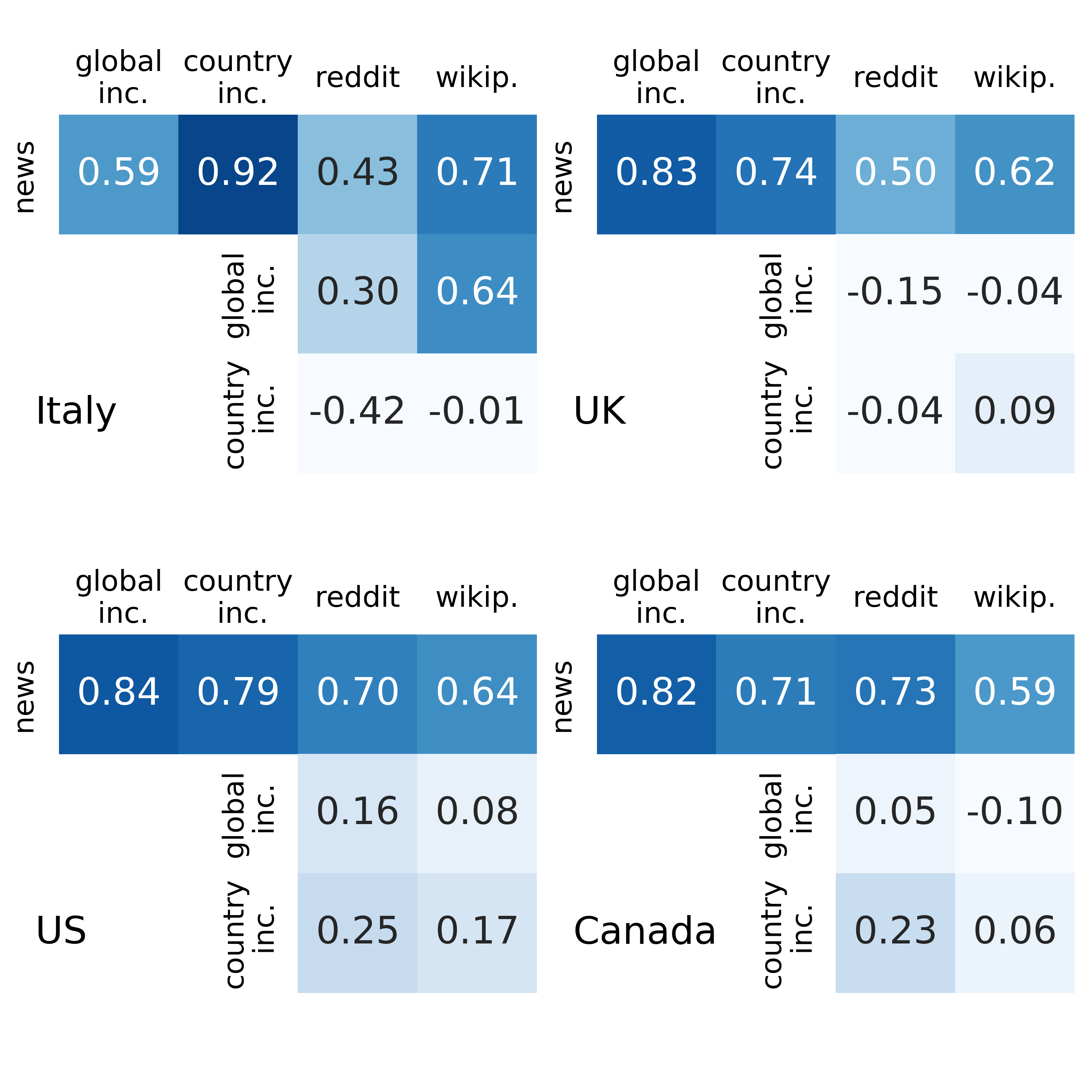}
\caption{\label{fig:correlation} 
Country specific Pearson correlation coefficients between: 1) news coverage and global/domestic COVID-19 incidence, volumes of Reddit comments and Wikipedia views; 2) global COVID-19 incidence and volumes of Reddit comments and Wikipedia views; 3) domestic COVID-19 incidence and volumes of Reddit comments and Wikipedia views.}
\end{figure}

The correlation between media coverage, public attention, and the epidemic progression is quantified more in details in Figure \ref{fig:correlation}. 
The plot shows that news coverage of each country is strongly correlated with COVID-19 incidence (both global and domestic), and slightly less with the volume of Reddit comments and Wikipedia views, which, in turn, are much less correlated with COVID-19 incidence (both global and domestic). This holds for all countries under consideration and highlights how the disease spreading triggers media coverage, and how the public response is more likely driven by such news exposure in each country rather than COVID-19 progression. 
Beyond these observations, it is interesting to notice from Figure \ref{fig:correlation} that Italy is the only country where news volume shows higher correlation with domestic rather than global incidence. This suggests that Italian media coverage follows more closely the internal evolution rather than the global one, at odds with respect to other countries. This is probably due to Italy being the location of the first COVID-19 outbreak outside Asia. 

\begin{figure}[tpb]%
    \centering
    \includegraphics[width=0.99\linewidth]{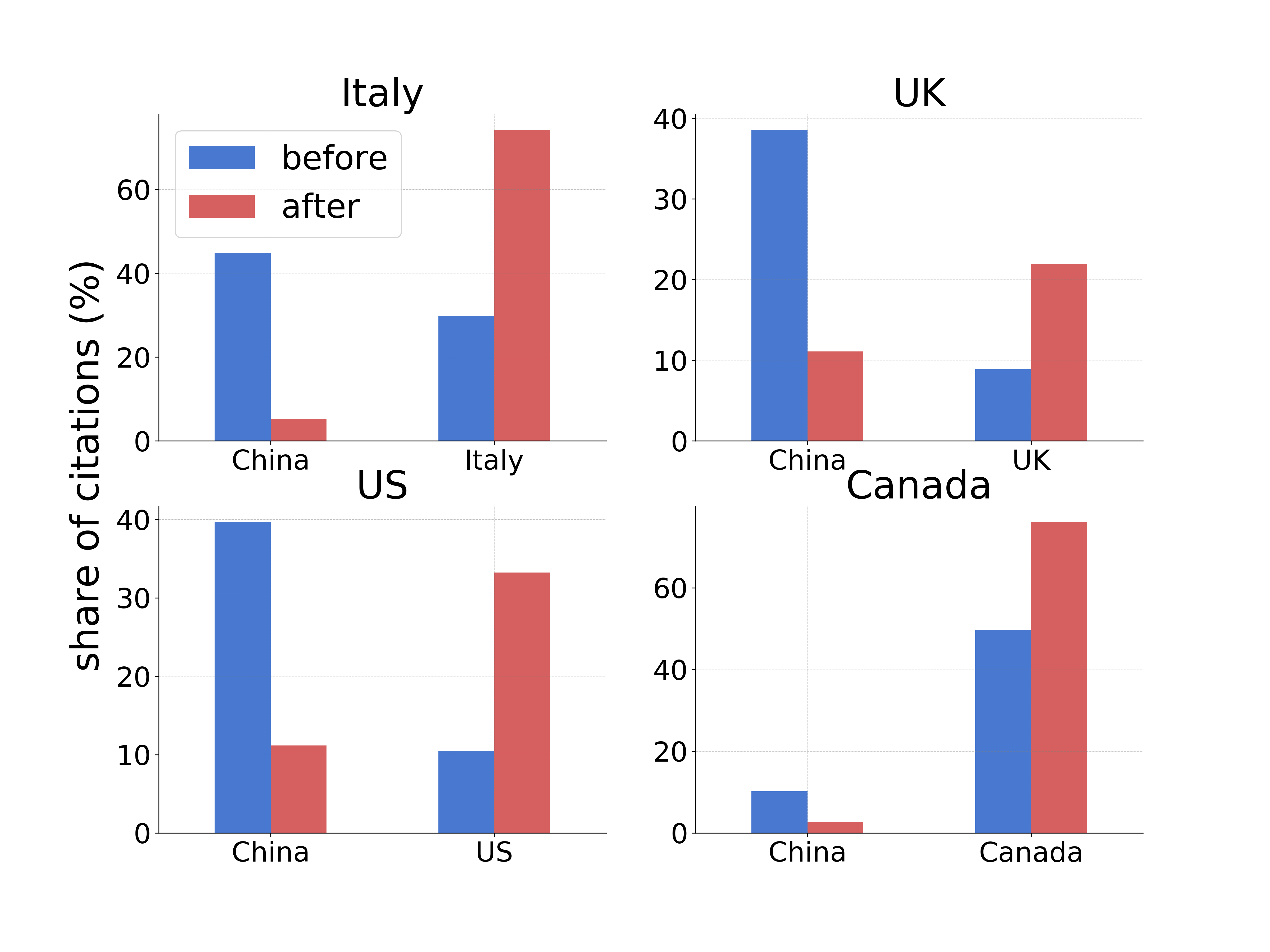}%
    \caption{Share of citations of China versus home country locations by Italian/UK/US/Canadian news outlets before and after first COVID-19 death occurred in different countries considered. Geographic locations are extracted from text using \cite{polyglot, geocoder}}.%
    \label{fig:geo}%
\end{figure}

This observation is supported by Figure \ref{fig:geo}, showing the citation share of Italian locations by Italian news media, before and after the first COVID-19 death was confirmed in Italy on 2020/02/20.  
After this date, Italian locations represent about $74\%$ of all places cited by Italian media (in our dataset), with an increase of $45\%$ with respect to the same statistics calculated before. 
Similar effects, though generally less intense, can be observed also in the other countries. 
Therefore, while media coverage is generally well synchronized with the global COVID-19 incidence, the media attention gradually shifts towards the internal evolution of the pandemic as soon as domestic outbreaks erupt. 
Arguably, this may have played an important role in individual risk perception. We can speculate that re-framing the emergency within a national dimension had the potential to amplify the perceived susceptibility of individuals \cite{ebolaperception, ebolamediamessages} and thus increase the adoption of behavioural changes \cite{hbm,gozzi2020towards}. 
Indeed, previous studies showed how at the beginning of February 2020 people were overly optimistic regarding the risks associated with the new virus circulating in Asia, and how their perception sharply changed after first cases were confirmed in their countries \cite{raude2020people, wise2020riskcovidusa}.

To explore more systematically the relationship between media coverage, public attention and epidemic progression, we consider a linear regression model to nowcast, separately for each country, collective public attention (quantified with the number of comments by geolocalized Reddit users or visits to relevant Wikipedia pages) given the volume of media coverage or the COVID-19 incidence as independent variables. We include also ``memory effects'' in the public attention by considering an exponential decaying term in the news time series \cite{tizzoni2020impact} (see subsection D of Methods and Materials for details). 
We compare the three models, where the independent variable(s) are the domestic incidence, the news volume, the news volume plus a memory term, by using the Akaike Information Criterion (AIC)~\cite{akaike1998information} and coefficient of determination ($R^{2}$). We found that the model considering only COVID-19 incidence has much less predictive power than the ones considering media coverage (see Table \ref{tab:fit} in Methods and Material). This enforces the idea that collective attention is mainly driven by media coverage rather than COVID-19 incidence. 
In addition, we found that including memory effects improves significantly the model performance. Not surprisingly, the coefficients of the ``memory effects'' term reported in  Table \ref{tab:coeff} are negative for all countries. This implies that public attention actually saturates in response to news exposure and gives us the chance to quantify the rate at which this phenomenon happens.

The results presented so far are in very good accordance with findings obtained in previous contexts related to epidemics and pandemics. Indeed, a similar media-driven spiky unfolding of public attention, measured through the information seeking and public discussions of online users, has been observed during the 2009 H1N1 influenza pandemic \cite{h1n1publicanxiety, h1n1twitter}, the 2016 Zika outbreak \cite{pruss2019zikatwitter}, the seasonal flu \cite{smith2015towardsRM} and during more localized public health emergency such as the 2013 measles outbreak in Netherlands \cite{mollema2015measles}. Our findings confirm the central role of media, showing how media exposure is capable of shaping and driving collective attention during a national and global health emergency.

Media exposure is an important factor that can influence individual risk perception as well \cite{mediariskperception, swinefluhype, mediacoveragemodel}. The timing and framing of the information disseminated by media can actually modulate the attention and ultimately the behaviour of individuals~\cite{funk2010review}. This becomes an even greater concern in a context where the most effective strategy to fight the spreading are containment measures based on individuals' behaviour. 
For this reason, in the next section we characterize 
media coverage and online users response more specifically in terms of content produced and consumed.

\begin{table}[tbp]
\begin{tabular}{l|cc|cc|cc}
\hline
       & \multicolumn{2}{c|}{\textit{news}} & \multicolumn{2}{c|}{\textit{$news_{MEM}$}} & \multicolumn{2}{c}{$R^{2}$} \\ \hline
       & reddit           & wikipedia          & reddit              & wikipedia              & reddit    & wikipedia   \\
Italy  & 0.87             & 0.43               & -0.41               & -0.15                  & 0.82      & 0.73        \\
UK     & 0.95             & 0.99               & -0.44               & -0.47                  & 0.82      & 0.85        \\
US     & 1.07             & 0.87               & -0.48               & -0.44                  & 0.88      & 0.82        \\
Canada & 1.12             & 1.06               & -0.40               & -0.45                  & 0.90      & 0.82        \\ \hline
\end{tabular}
\caption{\label{tab:coeff} Coefficient estimates for news plus memory effects linear regression model and related $R^{2}$. It is worth underlying that, despite the simplicity of this approach, we obtain relatively high values of $R^{2}$.}
\end{table}

\section{Dynamics of content production and consumption}
\label{sec:2}

While collective attention and media coverage are well correlated in terms of volume, the content and topics discussed by media and consumed by online users may not be as synchronized \cite{twitternews, burstytopics}. 
To shed light on this issue, we adopt an unsupervised topic modeling approach to extract prevalent topics in the news articles mentioned and discussed on Reddit. Often, indeed, users on Reddit post a submission containing a news article, and discussion unfolds in comments under such submission. Differently from the first part and to provide a comprehensive overview of the topics discussed, here we do not take into account any geographical context. 
Nonetheless, in the Supplementary Information we provide some insights also on the specific topics discussed by users in different countries.

We characterize the main topics discussed on Reddit by considering all submissions that include a news article in English.  
We then apply a topic modelling approach on the content of this news article set. 
Specifically, we extract topics by means of non-Negative Matrix Factorization (NMF) \cite{lee1999learning}, a popular method for this kind of tasks (see subsection E of Methods and Materials for details). In this way, we extract the $n=64$ most relevant topics in the news shared on Reddit.
As a second step, we apply the model trained on the Reddit news to the set of articles published by mainstream media. 
That is, we characterize the news published by media in terms of the topics discussed on Reddit. 
This choice allows us to directly compare the topics covered by media with the public discussion around such news exposure. A complete list of the $64$ topics extracted with the most frequent words is provided in the Supplementary Information.  

\begin{figure}[tpb]%
    \centering
    \label{fig:topic_scatter}
    \includegraphics[width=0.99\linewidth]{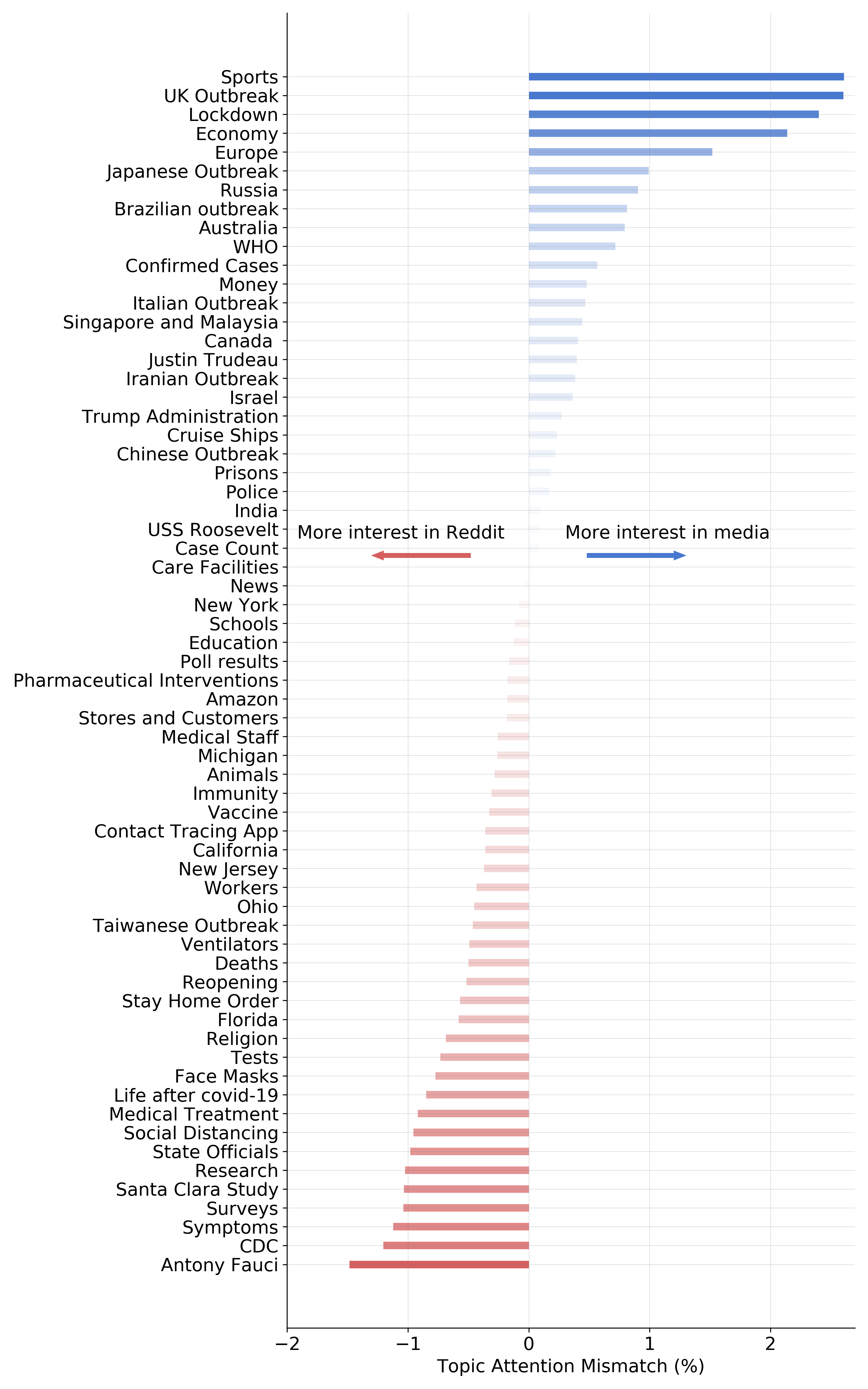}%
    \caption{\label{fig:topics} Difference in interest percentage share of different topics by traditional media and Reddit users. For example, $+2\%$ on the x-axis indicates that traditional media dedicates proportionally $2\%$ more attention to that specific topic with respect to Reddit users.}%
\end{figure}

We consider the number of articles published on a certain topic as a proxy of general interest of traditional media towards it, while we measure the collective interest of Reddit users by the number of comments under the news articles on a specific topic. 
Figure \ref{fig:topics} shows an overview of the topics extracted and a comparison of the interest of media and Reddit users.  
We find a diverse and heterogeneous set of topics. Among others, we recognize topics about the global spreading of the virus (Outbreaks, WHO, CDC), COVID-19 symptoms, treatment, hospitals and care facilities (Symptoms, Medical Treatment, Medical Staff, Care Facilities), the economic impact of the pandemic and responses from the governments to the upcoming crisis (Economy, Money), different societal aspects (Sports, Religious Services, Education), and also the possible interventions to mitigate the spreading of the virus (Face Masks, Social Distancing, Tests, Vaccine).

Overall, the attention of traditional media and Reddit users towards different topics are in good accordance.
Indeed, in Figure \ref{fig:topics} we represent the difference between interest share towards different topics in media and Reddit submissions. That is, we compute the percentage share of attention dedicated by news outlets and Reddit users to each topic, and we subtract these two quantities. We observe a maximum absolute mismatch in interest share of $2.61\%$. Nonetheless, we observe that Reddit users are slightly more interested to topics regarding health (Symptoms, Medical Treatment), non-pharmaceutical interventions and personal protective equipment (Social Distancing, Face Masks), studies and information on the epidemic (Research, Surveys, Santa Clara Study, CDC), and also to specific public figures such as Anthony Fauci.
Interestingly, the Santa Clara Study topic refers to the discussion about a controversial scientific paper suggesting that a much higher fraction of the population in the Santa Clara County was infected respect to what originally thought~\cite{Bendavid2020.04.14.20062463}. Since the study suggests a lower mortality rate, the preprint has been quickly leveraged to support protest against lockdowns\footnote{https://www.nytimes.com/2020/05/14/opinion/coronavirus-research-misinformation.html}, while substantial flaws have been detected in the scientific methodology of the paper~\footnote{https://www.theatlantic.com/health/archive/2020/04/pandemic-confusing-uncertainty/610819/}.

\begin{figure}[tpb]%
    \centering
    \includegraphics[width=0.99\linewidth]{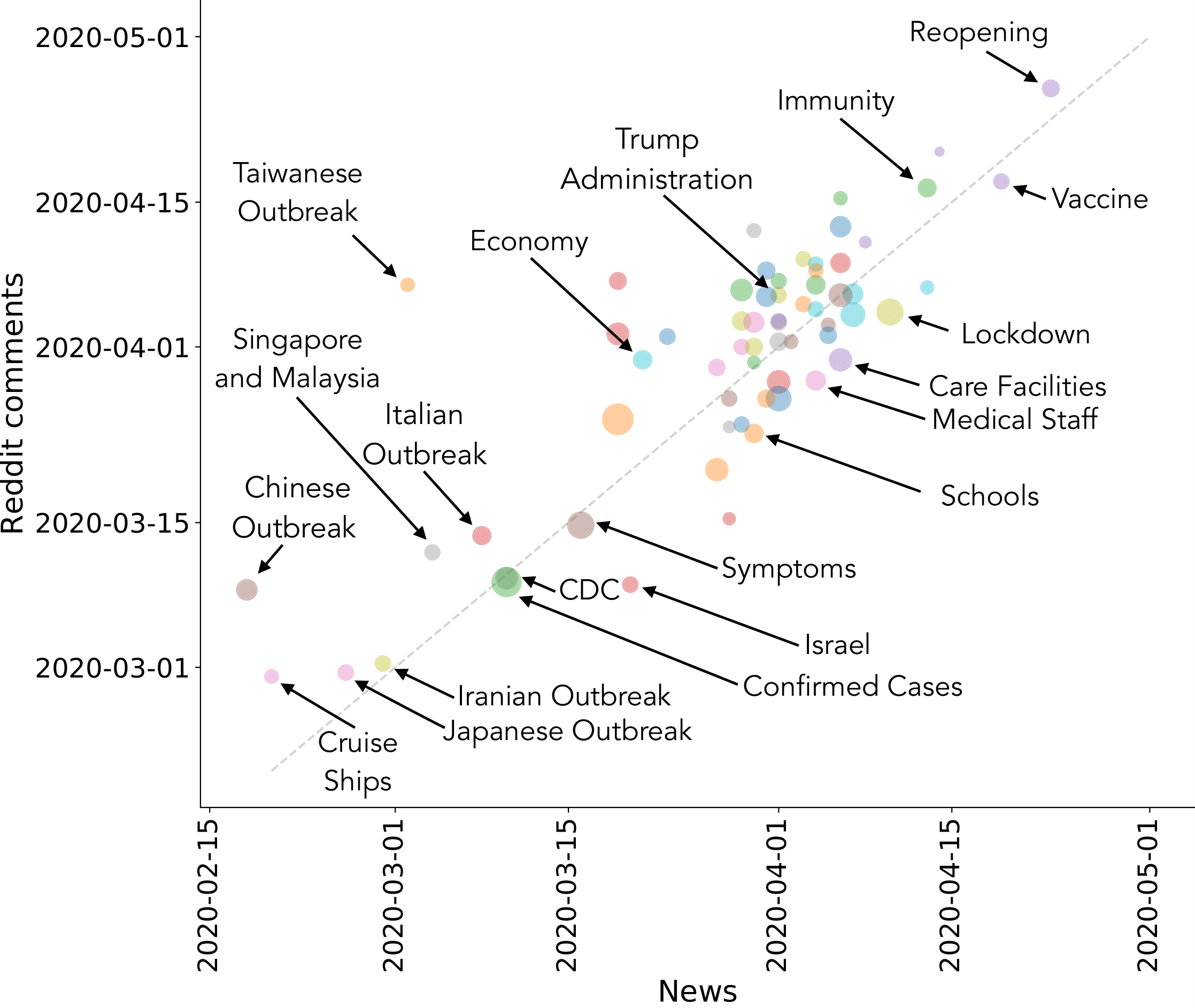}%
    \caption{\label{fig:scattertemporal} Scatter plot with the 64 topics extracted via NMF. X-axis (Y-axis) coordinate indicates when the topic achieved $50\%$ of its relevance in news outlets (Reddit) during our analysis interval.}
\end{figure}

The topics overview presented so far does not take into account any temporal dynamics of interest. However, topics showing a similar overall statistics may present a mismatch in temporal patterns. Hence, in the following, we take into account the temporal evolution of interest towards different topics. In Figure \ref{fig:scattertemporal} we represent each topic as a single point: its x-coordinate (y-coordinate) indicates when such topic reached $50\%$ of its total relevance in news outlets (on Reddit) during the analysis interval (see subsection E of Methods and Materials for a formal definition of the relevance of a topic).
Therefore, topics at the bottom left became relevant very early in the public discussion. 
Among these, we recognize themes centred on early COVID-19 outbreaks (i.e., Chinese, Japanese, Iranian and Italian outbreaks), the events related to cruise ships, specific countries (i.e., Israel, Singapore and Malaysia), and also topics regarding (early) health issues such as Symptoms, Confirmed Cases and the CDC. 

On the contrary, topics in the top right became relevant toward the end of the analysis interval (early May). 
Reasonably, we find here topics about the resumption of activities after lockdown (i.e. Reopening), the feasibility and timing of a possible vaccine against Sars-Cov-2 (i.e. Vaccine), and  discussions regarding acquired immunity and antibodies tests (i.e. Immunity). 
In-between, we find all other topics clustered around end of March and mid-April 2020, the period when the general discussion surrounding COVID-19 pandemic aroused sharply, as also shown in Figure \ref{fig:volume}.

Note that the diagonal (plotted as a dashed line) in Figure \ref{fig:scattertemporal} separates topics according to their temporal evolution. 
Above (below) the diagonal, we find topics whose interest on Reddit grows slowly (quickly) with respect to the media coverage. 
Therefore, above the diagonal the interest of Reddit users is mainly triggered by media exposure, while below it the interest grows faster and declines rapidly despite sustained media exposure. 
While the top-left and bottom-right regions are empty, indicating that, as a first approximation, temporal patterns of attention by traditional media and Reddit users are well-synchronized, interesting deviations from the diagonal are observable. 
For example, above the diagonal one can find mainly topics related to various outbreaks, economics and politics, for which the interests on Reddit follows the media coverage.
Below the diagonal, we observe topics more related to everyday life, such as Schools, Medical Staff, Care Facilities, and Lockdown, for which the attention on Reddit accelerates with respect to media coverage, and then declines rapidly.
Note that our view of topics discussed on Reddit is limited, since we only consider topics from news articles shared in submissions and do not explicitly take into account content expressed in comments. 
 This ensures a proper comparison with topics extracted from news published and  explains the absence of points in the bottom right corner of Figure~\ref{fig:scattertemporal}.

\section{Conclusions}

In this work, we characterized the response of online users to both  media coverage and COVID-19 pandemic progression. 
As a first step, we focused on the impact of media coverage on collective attention in different countries, characterized as volumes of country-specific Wikipedia pages views and comments of geolocalized Reddit users.
We showed that collective attention was mainly driven by media coverage rather than epidemic progression, 
rapidly saturated, and decreased despite media coverage and COVID-19 incidence remaining high. This trend is very similar to that observed during other outbreaks \cite{h1n1twitter, h1n1publicanxiety, pruss2019zikatwitter, smith2015towardsRM, mollema2015measles}.
 Also, we showed how media coverage sharply shifted to the domestic situation as soon as the first death was confirmed in the home country, discussing the implications for re-shaping individuals perception of risk \cite{raude2020people, wise2020riskcovidusa}.

As a second step, we focused on the dynamics of content production and consumption.
We modeled topics published in mainstream media and discussed on Reddit, showing that Reddit users were generally more interested in health, data regarding the new disease, and interventions needed to halt the spreading  with respect to media exposure. 
By taking into account the dynamics of topics extracted, we show that, while their temporal patterns are generally synchronized,
the public attention for topics related to politics and economics is mainly triggered by media exposure, while the interests for topics more related to daily life accelerates on Reddit with respect to media coverage.

Of course, our research comes with limitations. 
First, we characterized the exposure of individuals to COVID-19 pandemic by considering only news articles and Youtube videos published online by major news outlets. However, individuals are also exposed to relevant information through other channels, with television on top of these \cite{infoseekingTV}. 
Second, a $2013$ Pew Internet Study found that Reddit users are more likely young males \cite{duggan2013online}, showing that around $15\%$ of male internet users aged between $18$ and $29$ declare to use Reddit, compared to the $5\%$ of women in the same age range and to the $8\%$ of men aged between $30$ and $49$. 
Similarly, informal surveys proposed to users showed that most of respondents were males in their ``late teens to mid-20s'', and that female users were ``very much in the minority'' \cite{agegenderreddit}. Furthermore, Reddit is much more popular among urban and suburban residents rather than individuals living in rural areas \cite{duggan2013online}. 
Besides socio-demographic biases, other works suggested also that Reddit has become more and more a self-referential community, reinforcing the tendency to focus on its own contents rather than external sources \cite{redditselfcommunity}. Thus, perceptions, interests, and behaviours of Reddit users may differ from those of the general population. 
A similar argument may be raised for Wikipedia searches. Indeed, the usage of Internet, especially for information seeking purposes, can vary across people with different socio-demographic backgrounds \cite{digitaldivide, internetskills, digitalinequalities, VANDEURSEN2011125}. 

Finally, our view on online users reaction is partial. Indeed, we do not consider other popular digital data sources such as, for example, Twitter. The reason behind this choice is twofold. First, many studies already characterized public response during the current and past health emergencies through the lens of Twitter~\cite{park2017towards, lamb2012twitterpi,singh2020first,park2020covid19twitter,h1n1twitter,pruss2019zikatwitter,guidry2017ebola,martinez2018twitter,park2020conversations}.
Second, several studies have reported high prevalence of bots as drivers of low quality information and discussions on COVID-19 on this platform~\cite{ferrara2020covid,yang2020prevalence,singh2020first,ahmed2020covid,gallotti2020assessing}. Thus, careful and challenging extra steps would be necessary to isolate, identify, and distinguish organic discussions/reactions possibly originated from traditional media from those sparked by social bots. We leave this for future work.

In conclusion, our work offers further insights to interpret public response to the current global health emergency and raises questions about possible undesired effects of communication.
On one hand, our results confirm the pivotal role of media during health emergencies, showing how collective attention is mainly driven by media coverage. 
Therefore, since people are highly reactive to the news they are exposed to, in the beginning of an outbreak, the quality and type of information provided might have critical effects on risk perception, behaviours, and ultimately on the unfolding of the disease.
On the other hand, however, we found that collective online attention saturates and declines rapidly despite media exposure and disease circulation remaining high. 
Attention saturation has the potential to affect collective awareness, perceived risk and ultimately propensity towards virtuous individual behavioural changes aimed at mitigating the spreading. Furthermore, especially in case of unknown viruses, attention saturation might exacerbate the spreading of low quality information, which is likely to spread in the early phases of the outbreak when the characteristics of the disease are uncertain.  
Future works are needed to characterize the actual effects of attention saturation on human perceptions during a global health emergency.

Our findings suggest that public health authorities should consider to reinforce specific communication channels, such as social media platforms, in order to compensate the (natural) phenomenon of attention saturation. Indeed, these channels have the potential to create a more durable engagement with people, through a continuous loop of direct interactions. Currently, we see public health authorities issuing regularly declarations on social media. However, the CDC didn't even have a Twitter account in $2009$ during H1N1 pandemic (the account was created in May $2010$). While this is just an example, it underlines how we are relatively new to communicating such global health emergencies through social medias. Therefore, there is great need to further reinforce and engage people through these channels.
Alongside, public health authorities should consider to strengthen additional communication channels.
An example can be represented by participatory surveillance platforms all over the world such as Influenzanet, Flu Near You and FluTracking \cite{influenzanet, flunearyou, flutracking}, which have the potential of delivering in-depth targeted information to individuals during public health emergencies, to promote the exchange of information between people and public health authorities, with the potential to enhance the level of engagement in the community \cite{participatory}.

\section{Methods and Material}
\label{sec:mem}

In this section we provide general information about the data sets collected and the methods used. 

\subsection{News Articles and Videos}
We collect news articles using News API, a service that allows to download articles published online in a variety of countries and languages~\cite{newsapi}. For each of the country considered, we download all relevant articles published online by selected sources in the period 2020/02/07 - 2020/05/15. We select ``relevant'' articles considering those citing one of the following keywords: 'coronavirus', 'covid19', 'covid-19', 'ncov-19', 'sars-cov-2'. Note that for each article we have access to title, description and a preview of the whole text.
In total, our dataset consists in $227,768$ news: $71,461$ published by Italian, $63,799$ by UK, $82,630$ by US, and $9,878$ by Canadian media.

Additionally, we collect all videos published on YouTube by major news organizations, in the four countries under investigation, via their official YouTube channels using the official API~\cite{youtubeAPI}. In doing so, we download title and description of all videos and select as relevant those that mention one of the following keywords: `coronavirus', `virus', `covid', `covid19', `sars', `sars-cov-2', `sarscov2'. 

The reach of each channel (measured by number of subscribers) varies quite drastically from more than $9$ million for CNN (USA) to about $12$ thousand for Ansa (Italy). In total, the YouTube dataset consist of $13,448$ videos: $3,325$ by Italian, $3,525$ by British, $6,288$ by American, ans $310$ by Canadian channels. \\

It is important to underline that, while there is a good overlap between the sources of news articles and videos, some do not match. This is due to the fact that not all news organizations run a YouTube channel and others do not produce traditional articles. In the Supplementary Information, we provide a complete list of news outlets and Youtube channels considered.

\subsection{Reddit data set}

Reddit is a social content aggregation website where users can post, comment and vote content.
It is structured in sub-communities (i.e. \textit{subreddits}), centered around a variety of topics.

Reddit has already proven to be suitable for a variety of research purposes, ranging from the study of user engagement 
and interactions between highly related communities~\cite{tan2015all,hessel2016science} to post-election political analyses \cite{barthel_2016}. Also, it has been used to study the impact of linguistic differences in news titles  \cite{horne2017impact} and  to explore recent web-related issues such as hate speech \cite{saleem2017web} or cyberbullying \cite{rakib2018using} as well as health related issues like mental illness \cite{de2014mental}, also providing insights about the opioid epidemics~\cite{park2017towards,balsamo2019firsthand}.

We used the Reddit API to collect all submissions and comments published in Reddit under the subreddit /r/Coronavirus from 15/02/2020 to 15/05/2020.
After data cleanup by removing entries deleted by authors and moderators, we keep only submissions with score $> 1$ to avoid spam. We remove comments with less than $10$ characters and with more than $3$ duplicates, to avoid using automatic messages from moderation.  Final data contains $107,898$ submissions and $3,829,309$ comments from $417,541$ distinct users.

For the submissions, we then selected entries with links to English news outlets. The content of the urls was extracted using the available implementation\footnote{https://github.com/jcpeterson/openwebtext} of the method described in~\cite{radford2019language}, resulting in $66,575$ valid documents.

Reddit does not provide any explicit information about users' location, therefore we use self reporting via regular expression to assign a location to users. Reddit users often declare geographical information about themselves in submissions or comment texts.
We used the same approach as described in~\cite{balsamo2019firsthand}, that found the use of 
regular expressions as reliable, resulting in high correlation with census data in the US, although we acknowledge a potential higher bias at country level due to heterogeneities in Reddit population coverage and users demographics.
We selected all texts containing expressions such as `I am from' or `I live in'
and extracted candidate expressions from the text that follows the expression, to identify which ones represented country locations.
By removing inconsistent self reporting
we were able to assign a country to $789,909$ 
distinct users, from which $41,465$ have written at least one comment in the subreddit r/Coronavirus ($13,811$ from USA, $6,870$ from Canada, $3,932$ from UK and $445$ from Italy).

\subsection{Wikipedia data set}
Wikipedia has become a popular digital data source to study health information seeking behaviour \cite{h1n1publicanxiety, laurent2009wikiHISB}, and to monitor and forecast the spreading of infectious diseases \cite{brownstein2014wikipediaflu, generous2014wikidiseaseforecast}.
Here, we use the Wikimedia API \cite{wikiapi} to collect the number of visits per day of Wikipedia articles and the total monthly accesses to a specific project from each country. We consider the language as indicative of a specific country, suggesting the relevant projects for our analysis to be in English and Italian, i.e. \textit{en.wikipedia} and \textit{it.wikipedia} respectively.  We choose the articles directly related to COVID-19 and the ones in the \textit{'see also'} section of each page at the time of the analysis, 2020/02/07 - 2020/05/15, including country-specific articles (see Supplementary Information for full list of web pages considered). 

Except for the Italian, where the language is highly indicative of the location, the number of the access to English pages are almost evenly distributed among English-speaking countries. To normalize the signal related to each country we weight the number of daily accesses to a single article from a specific project $p$, $S_p(d)$, with the total number of monthly accesses from a country $c$ to the related Wikipedia project $T^{c}_p(d)$ such that the daily page views from a given Wikipedia project and country is 
\[y^{c}_{a, p}(d) = \frac{S_p(d) T^{c}_p(d)}{\sum_{c}T^{c}_p(d)},\] where the denominator is the total number of views of the Wikipedia specific project. The total volume of views at day $d$ from a country $c$ is then given by the sum over all the articles $a$ and projects $p$, namely
\[y^{c}(d) = \sum_{a,p}y^{c}_{a,p}(d).\]

 \subsection{Linear regression approach to model collective attention}
Above we showed how media coverage, COVID-19 incidence, and public attention are correlated even across four different countries. To move a step forward in this analysis, we considered a linear regression model that predicts for each country the public response given the news exposure. To include ``memory effects'' in the public response to media coverage, we consider also a modified version of this simple model, in which we weight cumulative news articles volume time series with an exponential decaying term~\cite{tizzoni2020impact}. Formally, we define the new variable:
\begin{equation}
    newsMEM = \sum_{\Delta t = 1}^{\tau} e^{-\frac{\Delta t}{\tau}} news(t - \Delta t)
\end{equation}
Where $\tau$ is a free parameter that sets the memory time scale and is tuned with cross-validation (more details in the Supplementary Information). 
These two models are compared to a linear regression that considers only COVID-19 incidence to predict public collective attention. Then, the models considered are: 
\begin{equation}
\begin{split}
\text{model I)   } y_{t}  & = \alpha_{1} incidence_{t} + u_{t} \\
\text{model II)  } y_{t}  & = \alpha_{1} news_{t} + u_{t} \\
\text{model III) } y_{t}  & = \alpha_{1} news_{t} + \alpha_{2} newsMEM_{t} + u_{t} \\
\end{split}
\end{equation}
Where $y_{t}$ can be either the volume of Reddit comments of geolocalized users or country specific Wikipedia visits, and $u_{t}$ is the error term.
In Table \ref{tab:fit} we report the results of the three regressions in terms of Akaike Information Criterion (AIC)~\cite{akaike1998information}. We observe that the model considering both news volume and memory effects is generally the better choice, while the model considering COVID-19 incidence only is the worst.

\begin{table}[tbp]
\begin{tabular}{l|cc|cc|cc}
\hline
       & \multicolumn{2}{c|}{\textit{incidence}}        & \multicolumn{2}{c|}{\textit{news}} & \multicolumn{2}{c}{\textit{news + $newsMEM$}} \\ \hline
       & reddit                    & wikipedia & reddit     & wikipedia    & reddit         & wikipedia         \\
Italy  & 3.71                      & -98.73    & -29.68     & -121.3       & -76.44         & -138.49            \\
UK     & 64.24                     & 65.85     & -16.03     & -20.46       & -51.97         & -65.24           \\
US     & 86.02                     & 61.85     & -11.76     & -14.67       & -48.39         & -44.93            \\
Canada & 84.53                     & 68.65     & -28.53     & -13.99       & -69.26         & -53.08            \\ \hline
\end{tabular}
\caption{\label{tab:fit} Akaike Information Criterion for the three linear regression models applied to predict Reddit comments and Wikipedia visits. As a practical rule, a model $i$ is preferred to model $j$ if $AIC(j) - AIC(i) \geq 3$ }
\end{table}

\subsection{Topic Modeling}
\label{sec:mem_topic}

Topic modeling has emerged as one of the
most effective methods for classifying, clustering, and retrieving
textual data, and has been the object of extensive investigation in the literature.
Many topic analysis frameworks are extensions of well known algorithms, considered as state-of-the-art for topic modeling.
\textit{Latent Dirichlet Allocation} (LDA)~\cite{blei2003latent} is the reference for probabilistic topic modeling. \textit{Nonnegative matrix factorization} (NMF)~\cite{lee1999learning} is the counterpart of LDA for the matrix factorization community.

Although there are many approaches to temporal and hierarchical topic modeling~\cite{blei2006dynamic,dou2013ht,gobbo2019topic}, we choose to apply NMF to the dataset, and then build time-varying intensities for each topic using the articles publication date. Starting from a dataset $\mathcal{D}$ containing the news articles shared in Reddit, we extract words and phrases with the methodology described in~\cite{mikolov2013distributed}, discarding terms with frequency below 10, to form a vocabulary $\mathcal{V}$ with around $60k$ terms. Each document is then represented as a vector of term counts, in a \textit{bag-of-words} approach. We apply TF-IDF normalization~\cite{jones1972statistical} and extract a total of $K = 64$ topics through NMF:
\begin{equation}
\setlength\abovedisplayskip{5pt}
\setlength\belowdisplayskip{5pt}
\min\limits_{W,H} {\Vert \textbf{X} - \textbf{W} \textbf{H} \Vert}^2_F \, ,
\label{nmf-minimization}
\end{equation}
where $\Vert  \Vert ^2_F$ is the Frobenius norm and $\textbf{X} \in \mathbb{R}^{|\mathcal{D}|\times |\mathcal{V}|}$ is the matrix resulting form TF-IDF normalization,
subject to the constraint that
the values in $\textbf{W} \in \mathbb{R}^{|\mathcal{D}|\times K}$ and $\textbf{H} \in \mathbb{R}^{K\times |\mathcal{V}|}$ must be nonnegative. 
The nonnegative factorization is achieved using the \textit{projected gradient method}
with sparseness constraints, as described in~\cite{lin2007projected,hoyer2004non}. The matrix $\textbf{H}$ is then used as a transformation basis for other datasets, e.g. with a new matrix $\widetilde{\textbf{X}}$ we fix $\textbf{H}$ and calculate a new $\widetilde{\textbf{W}}$ according to Eq.~\ref{nmf-minimization}.

For each topic $k$ we build a time series $\mathbf{s}_k$ for each dataset $\mathcal{D}$, where $s^{(t)}_{k}$ is the \textit{strength} of topic $k$ at time $t$. For the news outlets dataset, $s^{(t)}_{k} = \sum_{i \in \mathcal{D}^{(t)}}{w_{ik}}$, where $\mathcal{D}^{(t)}$ is the set of all documents shared at time $t$ in news outlets. For Reddit, we weight each shared document by its number of comments, and  $s^{(t)}_{k} = \sum_{i \in \mathcal{D}^{(t)}}{w_{ik}} \cdot c_i$, where $\mathcal{D}^{(t)}$ is the set of all documents shared at time $t$ in Reddit, and $c_i$ is the number of comments associated to document $i$. Finally, we define the \textit{relevance} of a topic as the integral in time of the strength. Therefore, given $t_{0}$ and $t_{f}$ as the start/end of our analysis interval, and given $R = \int_{t_{0}}^{t_{f}} dt s^{(t)}_{k}$, the coordinates of Figure \ref{fig:scattertemporal} are the $t_{1/2}$ such that $\int_{t_{0}}^{t_{1/2}} dt s^{(t)}_{k} = R/2$.

\begin{acknowledgments}
Authors would like to thank the startup Quick Algorithm for providing the platform \url{https://covid19.scops.ai/scops/home/}, where  the data collected during COVID-19 pandemic were visualized in real-time.
D.P. and M.T. acknowledge support from the Lagrange Project of the Institute for Scientific Interchange Foundation (ISI Foundation) funded by Fondazione Cassa di Risparmio di Torino (Fondazione CRT). M.T. acknowledges support  from EPIPOSE -``Epidemic intelligence to minimize COVID-19’s public health, societal and economical impact'' H2020-SC1-PHE-CORONAVIRUS-2020 call. M.S/ and A.P. acknowledge support from the Research Project ``Casa Nel Parco'' (POR FESR 14/20 - CANP - Cod. 320 - 16 - Piattaforma Tecnologica ``Salute e Benessere'') funded by Regione Piemonte in the context of the Regional Platform on Health and Wellbeing. A.P.  acknowledges  partial  support  from Intesa  Sanpaolo  Innovation  Center.
The  funders  had  no  role  in  study  design,  data collection and analysis, decision to publish, or preparation of the manuscript. N.G. acknowledges support from the Doctoral Training Alliance.
\end{acknowledgments}

\section*{Contributions}
\noindent
N.G, M.S., D.P., A.P. and N.P. conceptualized the study. N.G., N.P., A.P. and M.T. collected the data. N.G., A.P. and F.C. performed analyses. N.G., M.S. and N.P. wrote the initial draft of the manuscript. N.G. and A.P. provided visualization. All authors (N.G., N.P., D.P., M.S., A.P., M.T., F.C.) discussed the research design, reviewed, edited, and approved the manuscript.

\providecommand{\noopsort}[1]{}\providecommand{\singleletter}[1]{#1}

\clearpage
\onecolumngrid
\appendix
\section{Supplementary Information}

\begin{figure}[tpb]%
    \centering
    \includegraphics[width=0.7\textwidth]{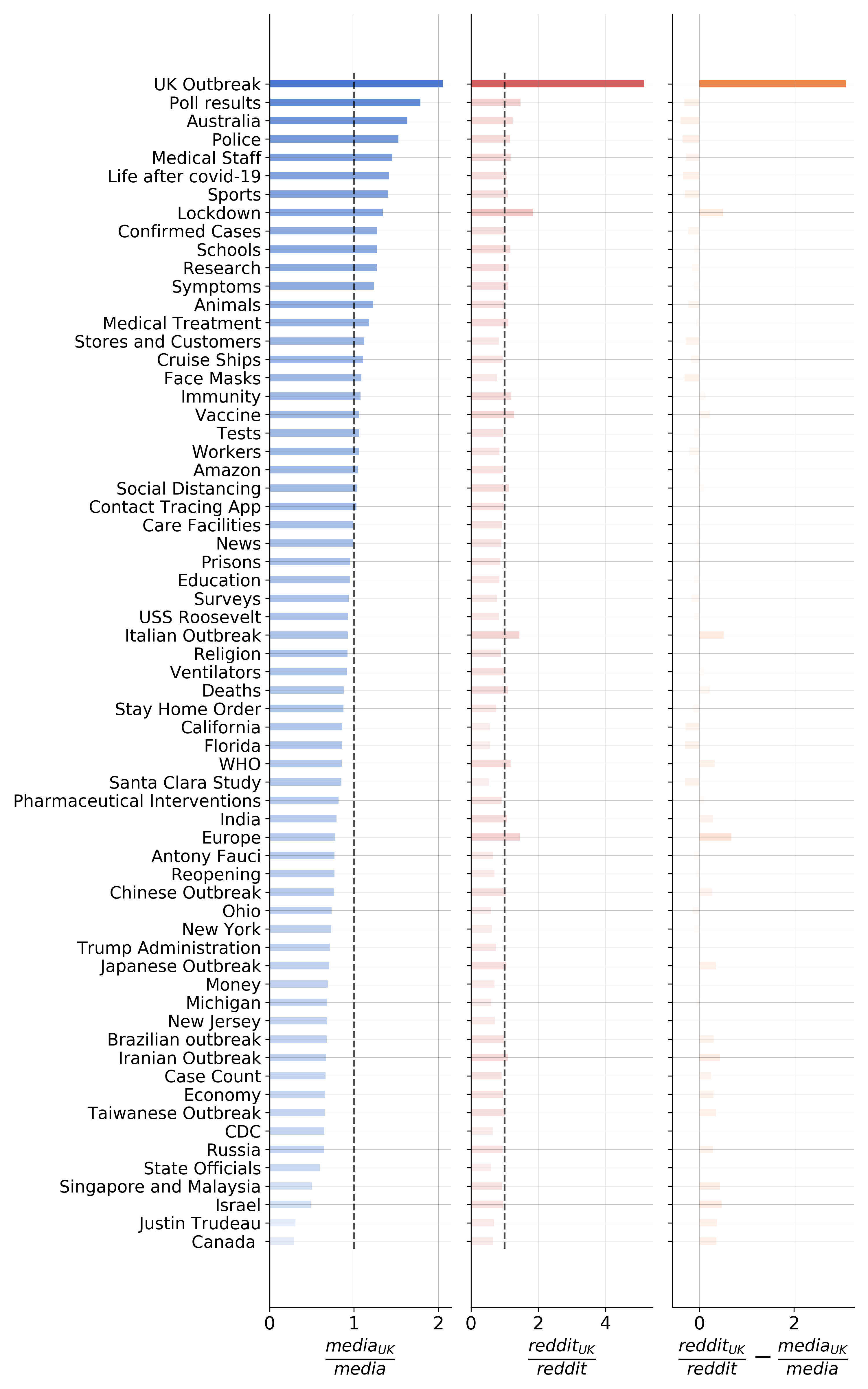}%
    \caption{\label{fig:topics_uk} From left to right: ratio between UK media interest and general media interest for different topics; ratio between UK Reddit users interest and general Reddit users interest for different topics; differences between these two quantities for different topics. In the fisrt two plots, topics to the left of the dashed line (on $1$) are less discussed in by UK media/users with respect to the general discussion, while topics to the right are more discussed. In the last plot, positive (negative) bars indicated that UK Reddit users pay generally more (less) attention to that topic with respect to UK media.}%
\end{figure}

\begin{figure}[tpb]%
    \centering
    \includegraphics[width=0.7\textwidth]{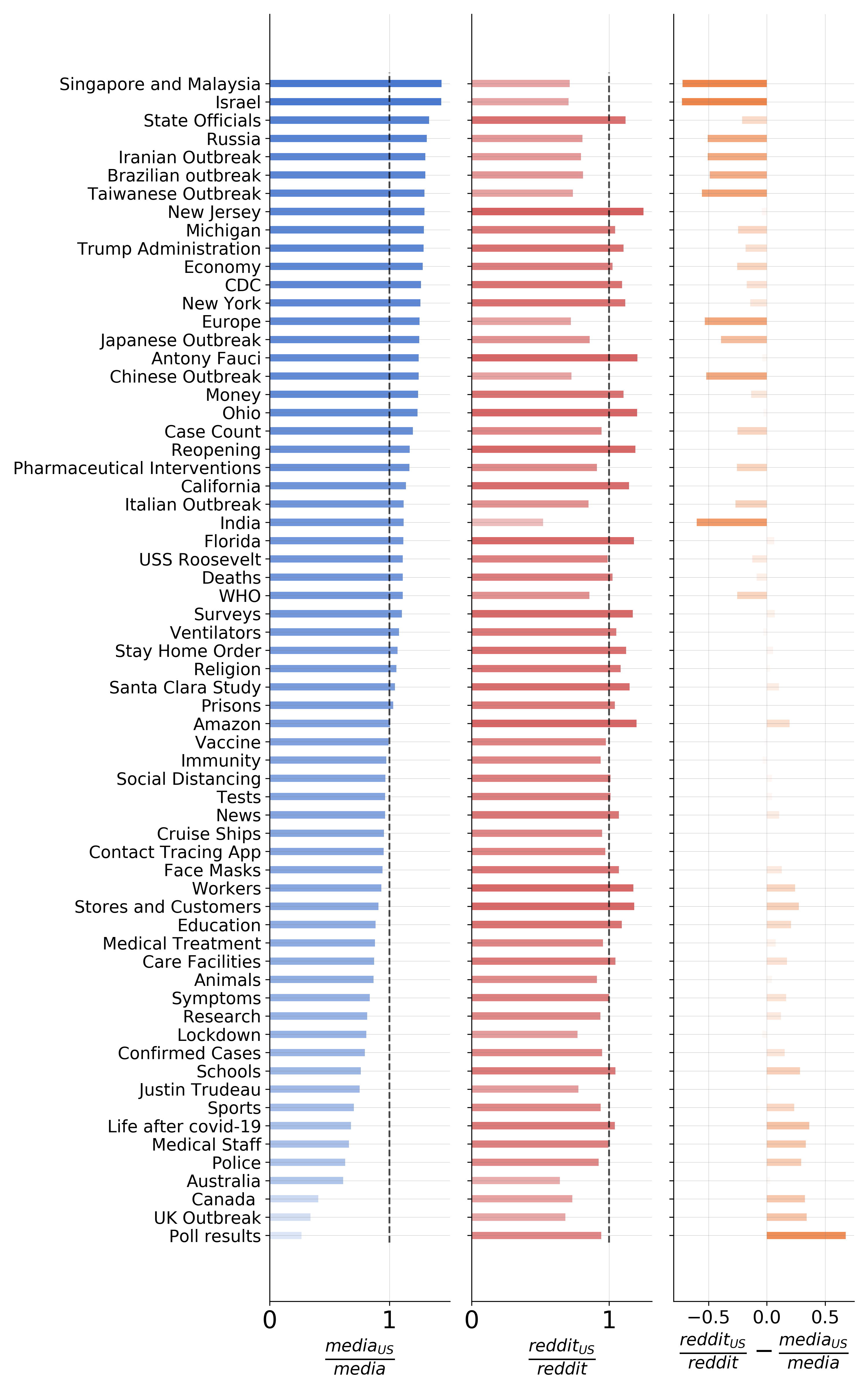}%
    \caption{\label{fig:topics_us} From left to right: ratio between US media interest and general media interest for different topics; ratio between US Reddit users interest and general Reddit users interest for different topics; differences between these two quantities for different topics. In the fisrt two plots, topics to the left of the dashed line (on $1$) are less discussed in by US media/users with respect to the general discussion, while topics to the right are more discussed. In the last plot, positive (negative) bars indicated that US Reddit users pay generally more (less) attention to that topic with respect to US media.}%
\end{figure}

\begin{figure}[tpb]%
    \centering
    \includegraphics[width=0.7\textwidth]{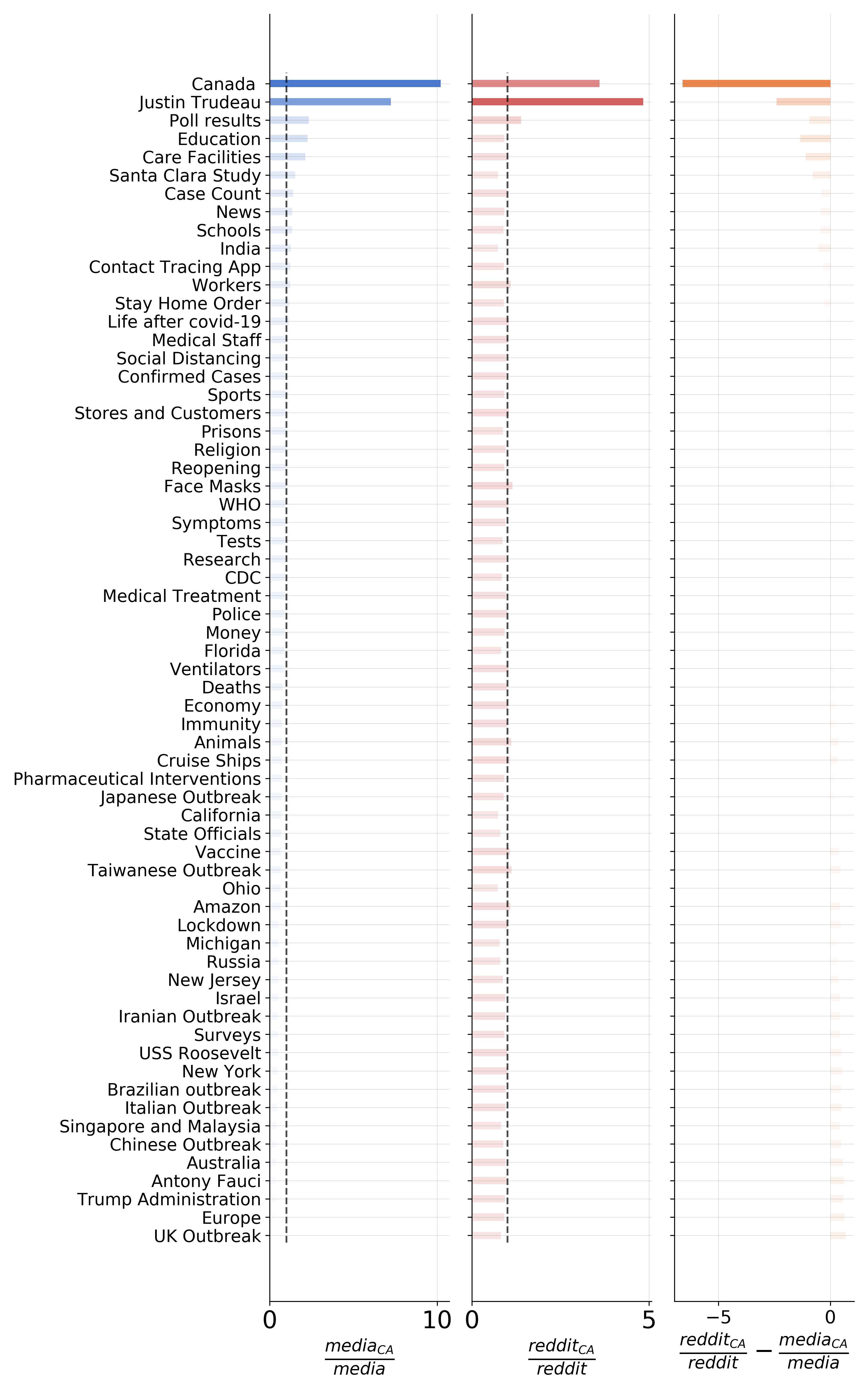}%
    \caption{\label{fig:topics_ca} From left to right: ratio between Canadian media interest and general media interest for different topics; ratio between Canadian Reddit users interest and general Reddit users interest for different topics; differences between these two quantities for different topics. In the fisrt two plots, topics to the left of the dashed line (on $1$) are less discussed in by Canadian media/users with respect to the general discussion, while topics to the right are more discussed. In the last plot, positive (negative) bars indicated that Canadian Reddit users pay generally more (less) attention to that topic with respect to Canadian media.}%
\end{figure}

\begin{figure}[tpb]%
    \centering
    \includegraphics[width=0.7\textwidth]{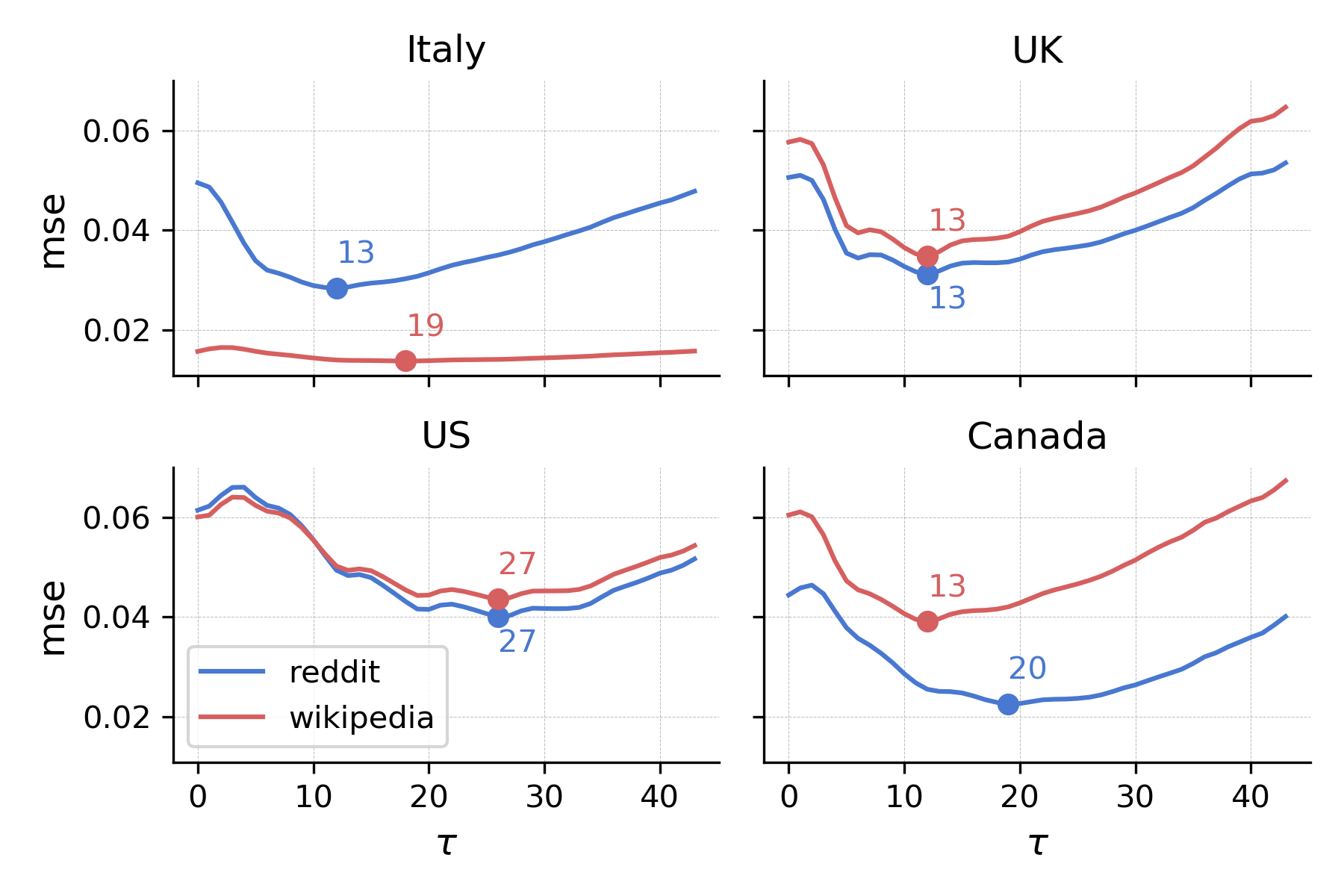}%
    \caption{\label{fig:tau} Cross validation scores (computed as mean-squared-error) as a function of $\tau$. For the four countries, we show the best value of $\tau$ both for Reddit comments and Wikipedia page views.}%
\end{figure}

In this Supplementary Information we provide the list of news sources (Table \ref{tab:newssources}), YouTube channels (Table \ref{tab:youtube}), Wikipedia pages (Table \ref{tab:wikiarticles}), and the $64$ topics extracted through NMF and the most frequent words for each one (Table \ref{tab:topics}). We also provide an insight on the topics discussed in different countries (Figures \ref{fig:topics_uk}, \ref{fig:topics_us}, \ref{fig:topics_ca}) and more details on the results of the cross-validation of the memory time scale parameter $\tau$ used in the equal-time regression model (Figure \ref{fig:tau}).
\begin{table*}[!ht]
\caption{Lists of news sources considered for Italy, United Kingdom, United States, and Canada. These lists aim to provide the most complete overview of the communication medias landscape in different countries, while facing with the limitation imposed by the API.
}
\begin{ruledtabular}
\begin{tabular}{llll}
\textrm{Italy}&
\textrm{United Kingdom}&
\textrm{United States}&
\textrm{Canada}\\
\colrule
ilfattoquotidiano.it & mirror.co.uk & cnbc & financial-post \\
rainews.it & express.co.uk & reuters & google-news-ca \\
ansa & standard.co.uk & nbc-news & the-globe-and-mail \\
tg24.sky.it & business-insider-uk & fox-news & cbc-news \\
corrieredellosport.it & bbc-news & bloomberg & \\
google-news-it & telegraph.co.uk & the-wall-street-journal & \\
la-repubblica & google-news-uk & cbs-news & \\
huffingtonpost.it & dailymail.co.uk & time & \\
ilgiornale.it & metro.co.uk & cnn & \\
il-sole-24-ore & independent & abc-news & \\
lastampa.it & thetimes.co.uk & & \\
wired.it & thesun.co.uk & & \\
corriere.it & & & \\
ilpost.it & & & \\
liberoquotidiano.it & & & \\
ilmessaggero.it & & & \\
\end{tabular}
\end{ruledtabular}
\label{tab:newssources}
\end{table*}

\begin{table*}[h]
\caption{Lists of YouTube channels considered for Italy, United Kingdom, United States, and Canada. In parentheses, we report the number of subscribers as of May $18th$.}
\begin{ruledtabular}
\begin{tabular}{llll}
\textrm{Italy}&
\textrm{United Kingdom}&
\textrm{United States}&
\textrm{Canada}\\
\colrule
antefattoblog (290,000) & BBCNews (7,540,000) & CNN (9,630,000) & cbcnews (2090000) \\
euronewsit (227,000) & skynews (2,340,000) & ABCNews (8,850,000) & GlobalToronto (1,670,000) \\
corrieredellasera (55,000) & Channel4News (1,330,000) &  FoxNewsChannel (5080000) & TheGlobeandMail (77,400) \\
lastampait (52,400) & telegraphtv (1,170,000) &  NBCNews (3080000)& thefinancialpost (15,500) \\
ilsole24ore (25,800) & thesunnewspaper (799,000) & businessinsider (3,060,000) & \\
ANSA (12,500) & ITVNews (530,000) & TheNewYorkTimes (2,860,000) & \\
&EveningStandardNews (74100) & CBSNewsOnline (2,640,000)&\\
& theindependent (46800)& WSJDigitalNetwork (2,160,000)&\\
& timesonlinevideo (41500)& cnbc (1,590,000)&\\
& dailymirror (23400)& USATODAY (1,460,000)&\\
& & HuffingtonPost (603,000)&\\
& & ReutersVideo (334,000)&\\
\end{tabular}
\end{ruledtabular}
\label{tab:youtube}
\end{table*}

\begin{table*}[h]
\caption{List of Wikipedia articles related to the COVID-19}
\begin{ruledtabular}
\begin{tabular}{lll|l}
\multicolumn{3}{c|}{en.wikipedia}&
it.wikipedia\\
\hline
\textrm{United Kingdom}&
\textrm{United States}&
\textrm{Canada}&
\textrm{Italy}\\
\hline
\colrule
\hline
\multirow{2}{*}{}COVID-19 pandemic in the & COVID-19 pandemic in the  & COVID-19 pandemic  & Pandemia di COVID-19 del 2020 \\ United Kingdom& United States & in Canada &  in Italia \\
\hline
\multicolumn{3}{l|}{Severe acute respiratory syndrome coronavirus 2} & SARS-CoV-2\\
\multicolumn{3}{l|}{COVID-19 pandemic lockdowns} &\\
\multicolumn{3}{l|}{Portal:Coronavirus disease 2019}	&	\\
\multicolumn{3}{l|}{COVID-19 vaccine} &				\\
\multicolumn{3}{l|}{COVID-19 drug development} &\\	
\multicolumn{3}{l|}{2019-2020 coronavirus pandemic} &	Pandemia di COVID-19 del 2019-2020\\
\multicolumn{3}{l|}{Coronavirus disease 2019} &							\\
\multicolumn{3}{l|}{COVID-19 pandemic} &			\\
\multicolumn{3}{l|}{Novel coronavirus} &					\\
\multicolumn{3}{l|}{COVID-19 in pregnancy} &	COVID-19 in gravidanza\\
\multicolumn{3}{l|}{Evacuations related to the COVID-19 pandemic} &		\\
\multicolumn{3}{l|}{\multirow{2}{*}{COVID-19 pandemic by country and territory}} & Pandemia di COVID-19 \\&&& del 2019-2020 nel mondo\\		
\multicolumn{3}{l|}{COVID-19 drug repurposing research} &	\\
\multicolumn{3}{l|}{List of deaths due to coronavirus disease 2019} & \\
\multicolumn{3}{l|}{List of events affected by the COVID-19 pandemic} & \\
\multicolumn{3}{l|}{Coronavirus disease}&	\\
\multicolumn{3}{l|}{\multirow{2}{*}{Travel restrictions related to the COVID-19 pandemic}}	& Restrizioni agli spostamenti correlate alla \\&&& pandemia di COVID-19 del 2019-2020\\
\end{tabular}
\end{ruledtabular}
\label{tab:wikiarticles}
\end{table*}

\begin{table*}[]
\caption{\label{tab:topics} List of $64$ topics extracted using NMF with most frequent words.}
\begin{tabular}{ll}
\hline
\textbf{topic name}          & \textbf{frequent words}                                                         \\ \hline
WHO                          & countries, world, pandemic, global, africa, health\_organization                     \\
Case Count                   & cases, new, reported, total, number, confirmed\_cases                                \\
State Officials              & state, officials, department, governor, washington, oregon                           \\
Medical Treatment            & patients, patient, hospitals, doctors, treatment, severe                             \\
Reopening                    & reopen, reopening, states, businesses, open, texas                                   \\
Chinese Outbreak             & china, chinese, wuhan, beijing, outbreak, epidemic                                   \\
Medical Staff                & hospital, hospitals, staff, nurses, doctors, medical                                 \\
Education                    & students, campus, university, student, classes, college                              \\
Trump Administration         & trump, white\_house, president, president\_donald, americans, administration         \\
Pharmaceutical Interventions & drug, remdesivir, hydroxychloroquine, drugs, treatment, trial                        \\
Tests                        & testing, tests, test, tested, labs, positive                                         \\
New York                     & new\_york, cuomo, city, blasio, mayor, gov\_andrew                                   \\
Stay Home Order              & order, stay\_home, governor, businesses, issued, services                            \\
Canada                       & province, ontario, quebec, henry, toronto, provincial                                \\
Vaccine                      & vaccine, vaccines, trial, moderna, trials, developed                                 \\
India                        & india, indian, delhi, new\_delhi, mumbai, maharashtra                                \\
Cruise Ships                 & passengers, ship, cruise\_ship, cruise, board, quarantine                            \\
Face Masks                   & masks, mask, face\_masks, equipment, use, n95\_masks                                 \\
Santa Clara Study            & county, officials, department, santa\_clara, public\_health, los\_angeles            \\
Life after covid-19          & time, family, home, want, right, things                                              \\
Deaths                       & deaths, died, reported, death, total, death\_toll                                    \\
UK Outbreak                  & uk, nhs, london, england, scotland, boris\_johnson                                   \\
Workers                      & workers, employees, work, company, employee, plant                                   \\
Italian Outbreak             & italy, italian, country, lombardy, rome, milan                                       \\
Poll results                 & cent, email\_address, mr, headlines, enter, men                                      \\
Florida                      & florida, desantis, gov\_ron, tampa\_bay, beaches, state                              \\
Japanese Outbreak            & japan, tokyo, japanese, abe, south\_korea, infections                                \\
CDC                          & cdc, public\_health, officials, disease\_control, centers, agency                    \\
Iranian Outbreak             & iran, iranian, tehran, qom, sanctions, outbreak                                      \\
Economy                      & economy, year, million, economic, companies, market                                  \\
Social Distancing            & population, social\_distancing, infected, number, disease, model                     \\
Schools                      & schools, school, children, parents, education, students                              \\
Surveys                      & percent, americans, survey, study, respondents, poll                                 \\
Israel                       & ministry, israel, quarantine, israeli, israelis, netanyahu                           \\
Michigan                     & michigan, whitmer, detroit, state, gov\_gretchen, protesters                         \\
Australia                    & australia, new\_zealand, australian, nsw, australians, ardern                        \\
Stores and Customers         & stores, customers, store, food, company, products                                    \\
Ventilators                  & ventilators, ventilator, equipment, hospitals, production, need                      \\
Prisons                      & inmates, prison, prisons, prisoners, jail, release                                   \\
Money                        & billion, money, house, program, help, congress                                       \\
California                   & newsom, california, los\_angeles, gov\_gavin, san\_francisco, state                  \\
Taiwanese Outbreak           & taiwan, taiwanese, taipei, island, hong\_kong, epidemic\_command                     \\
Amazon                       & amazon, company, warehouses, employees, bezos, warehouse                             \\
Police                       & police, officers, arrested, man, city, video                                         \\
Care Facilities              & residents, care, facility, nursing\_homes, nursing\_home, facilities                 \\
Contact Tracing App          & data, information, app, google, users, privacy                                       \\
Justin Trudeau               & canada, trudeau, canadians, canadian, ottawa, justin\_trudeau                        \\
Singapore and Malaysia       & singapore, moh, malaysia, linked, dormitories, related\_story                        \\
Religion                     & church, spell, services, pastor, churches, service                                   \\
Antony Fauci                 & fauci, dr\_anthony, allergy, national\_institute, infectious\_diseases, white\_house \\
Research                     & sars\_cov, study, researchers, research, scientists, viruses                         \\
Sports                       & games, players, fans, league, season, game                                           \\
Immunity                     & antibodies, blood, immunity, immune, recovered, antibody\_tests                      \\
News                         & news, weekday\_mornings, stories, coverage, nbc\_news, breaking\_news                \\
USS Roosevelt                & navy, sailors, ship, crozier, modly, guam                                            \\
Symptoms                     & symptoms, person, sick, spread, soap, water                                          \\
Europe                       & germany, france, spain, french, eu, german                                           \\
Ohio                         & ohio, dewine, acton, ohio\_department, ohioans, election                             \\
Lockdown                     & government, lockdown, country, measures, prime\_minister, minister                   \\
Brazilian outbreak           & bolsonaro, brazil, rio, brazilian, sao\_paulo, president\_jair                       \\
Russia                       & russia, moscow, putin, russian, kremlin, city                                        \\
New Jersey                   & murphy, new\_jersey, gov\_phil, nj\_com, persichilli, state                          \\
Confirmed Cases              & man, case, year\_old, woman, patient, confirmed                                      \\
Animals                      & animals, pets, animal, cats, hong\_kong, zoo                                         \\ \hline
\end{tabular}
\end{table*}

\end{document}